\documentstyle[prl,preprint,aps,epsf]{revtex}

\begin{document}

\draft
\title{Stochastic processes with finite correlation time: modeling and 
application to the generalized Langevin equation}

\author
{T. Srokowski}

\address{
 Institute of Nuclear Physics, PL -- 31-342 Krak\'ow, Poland }

\date{\today}

\maketitle

\begin{abstract}
\parbox{14cm}{\rm
The kangaroo process (KP) is characterized by various forms of the covariance 
and can serve as a useful model of random noises. We discuss properties of that
process for the exponential, stretched exponential and algebraic (power-law)
covariances. Then we apply the KP as a model of noise in the generalized
Langevin equation and simulate solutions by a Monte Carlo method. Some results
appear to be incompatible with requirements of the fluctuation-dissipation 
theorem because probability distributions change when the process is inserted 
into the equation. We demonstrate how one can construct a model of noise free 
of that difficulty. This form of the KP is especially suitable for physical
applications. 
\\}
\end{abstract}


\vfill
\newpage

\narrowtext

\section{Introduction}

It is frequently assumed that a noise in stochastic equations is not
correlated and the underlying stochastic process 
can be regarded as Markovian. This
assumption is justified only if the time scales involved are large compared to
the noise correlation time. On the other hand, the noise itself can result from
a procedure of fast modes removal. It is well known \cite{med,yak} that in 
such case noise correlations, both in space and time, must arise. For some
stochastic processes the noise covariance decays fast with time and 
it can be put in an exponential form (the coloured noise \cite{han}).
Frequently even longer tails of the noise covariance are observed. 
Algebraic covariances appear in the fluid 
dynamics \cite{maz,cho,ald} and linearized hydrodynamics \cite{cic}; 
they are responsible for such phenomena as noise-induced
Stark broadening \cite{fri1} and anomalous nuclear scattering 
in the framework of the molecular dynamics \cite{sro1}. A direct consequence of
the algebraic form of the velocity autocorrelation function, falling not 
faster then $1/t$, is the infinite value of the diffusion 
coefficient \cite{bou}. An anomalous diffusion
process is frequently observed in disordered media where a trapping mechanism
leads to algebraic distributions of waiting time. For example, some amorphous
insulating materials (e.g. As$_2$Se$_3$) exhibit a highly dispersive transient
photocurrent \cite{scharfe}, just due to charge hopping between spatially
disordered sites. 

Stochastic dynamics driven by a noise different from the white noise obeys 
the generalized Langevin equation (GLE) \cite{mori,lee}:
\begin{eqnarray}
\label{gle}
m \frac{dv(t)}{dt} = -m\int_0^t K(t-\tau)v(\tau)d\tau + F(t)~~~~\ 
\end{eqnarray}
where $F(t)$ is a stochastic force and $m$ denotes the mass of the Brownian
particle. Due to the second fluctuation-dissipation theorem (FDT) 
\cite{kubo,berne}, the kernel $K(t)$ can be expressed in terms of the noise 
covariance $C_F\equiv\langle F(0)F(t)\rangle$: $K(t)=C_F/mT, $ with the
temperature $T$. The Eq.(\ref{gle}) can be handled 
as a usual Volterra equation \cite{smirnov}. Assuming the initial condition
$v(0)=0$, the general solution can be expressed in the form of a stochastic
integral \cite{adel}:
\begin{equation}
\label{solv}
v(t)=m^{-1}\int_0^t R(t-\tau)\,F(\tau)\,d\tau,
\end{equation}
where the Laplace transform of the resolvent $R(t)$ is given by the equation 
\begin{equation}
\label{rods}
\widetilde R(s)=1/[s+\widetilde K(s)]. 
\end{equation}
From the Eq.(\ref{solv}), expressions for some average quantities follow. 
For the velocity variance, we have
\begin{equation}
\label{v2a}
\langle v^2\rangle_{S\;} (t)=m^{-2}\int_0^t\int_0^t R(t-\tau)\,R(t-\tau')\,
C_F(|\tau-\tau'|)\,d\tau d\tau',
\end{equation}
where the average $\langle \;\rangle_{S\;}$ is taken
over an equilibrium ensemble with some stationary probability distribution. 
The FDT ensures that asymptotically, for large time,
the system reaches the equilibrium value $\langle v^2\rangle_{S\;}=T/m$ 
(the equipartition energy rule). For the velocity autocorrelation 
function, in turn, we have simply:
\begin{equation}
\label{codt}
C_v(t)\equiv \langle v(0)v(t)\rangle_{S\;}=T/m\; R(t).
\end{equation}

The assumption about a form of noise covariance is sufficient to calculate some
average quantities. In order simulate stochastic trajectories from 
the Eq.(\ref{solv}), one needs a concrete physical process
which could serve as a model of the noise. For example, for the exponential 
covariance it could be the well-known Ornstein-Uhlenbeck process, which, due to
the Doob theorem \cite{doob}, is very important if amplitude
distributions are Gaussian. A broad 
class of stochastic processes known as "kangaroo processes" (KP) \cite{fri} is
especially interesting. One can construct the KP for an arbitrary, given form 
of covariance. The KP is particularly well suited for problems involving
algebraic, scale-invariant dependences. Long tails of constant value of the
process in the step-wise structure of the KP make possible to preserve 
the memory about this value for a sufficiently long time to produce 
such slowly decaying form of the covariance. Due to that
structure, the KP resembles stochastic, dispersive transport processes 
in disordered media, e.g. the hopping time distribution \cite{scher,nool,bos}.
In the framework of random walk processes, a pattern of long straight-line 
segments is typical for L\'evy flights \cite{kla,bar,bar1}.

This paper deals with random noises possessing various covariances: 
exponential, stretched exponential and algebraic, and expresses them in terms
of the KP. The most important properties of KP are summarized in
Sec.II; we also derive there formulas referring to those forms of
the covariance. In Sec.III we consider the application of the KP
as a model of random force in the GLE. The most important results are
summarized and discussed in Sec.IV. 

\section{The kangaroo process}

The kangaroo process \cite{fri} is a step-wise, discontinues random function. 
The value of the process, $m(t)$, is determined at subsequent random jumping 
times $t_1$, $t_2$, $\dots$. The jumping frequency $\nu(m)$ depends on the 
value of the process itself and $m$ remains constant between jumps. We 
introduce also the interval length as a reciprocal of the frequency: 
$s=1/\nu$. Due to some physical applications, this quantity can also be called 
"a free path". The KP is a stationary Markov process and can be 
defined by the probability density $p(m,t)$ satisfying the following 
Fokker-Planck equation 
\begin{equation}
\label{fpkp}
\frac{\partial}{\partial t}p(m,t) = \nu (m) \left(-p(m,t) + 
\frac{P_{KP}(m)}{\int \nu (m') P_{KP}(m') dm'} 
\int \nu (m') p(m',t) dm'\right) 
\end{equation}
where $P_{KP}(m)$ denotes a stationary probability distribution of $m(t)$. 
The interval length $s$ is also a stochastic quantity. Its probability
distribution $P(s)$ is connected with $P_{KP}(m)$ by the relation
$P(s)ds = 2 P_{KP}(|m|)d|m|$.
We assume that $P_{KP}(m)$ and $\nu(m)$ are even functions of $m$. This 
assumption allows us to get a simple expression for the covariance of KP
$C(t)=\langle m(t)m(0)\rangle_{S\;}$ where the average is taken over 
the stationary probability distribution $P_{KP}(m)$:
\begin{equation}
\label{f10}
C(t) = \int_{-\infty}^{+\infty} m^2 P_{KP}(m) \exp(-\nu(m) |t|) dm.
\end{equation}
We want to derive an expression for $\nu(m)$ for a given covariance $C(t)$ and
an amplitude distribution $P_{KP}(m)$. Let us assume that $\nu(m)$ is 
a monotonic increasing function of $|m|$ and $\nu(\infty) = \infty$.
Then we can change the integration variable
in Eq.(\ref{f10}) and obtain the Laplace integral
\begin{equation}
\label{f11}
C(t) = 2 \int_{\nu(0)}^{+\infty} m^2 P_{KP}(m) \frac{dm}{d\nu} 
\exp(-\nu |t|) d\nu.
\end{equation}
Therefore $\nu(m)$ is a solution of the following differential equation
\begin{equation}
\label{f12}
\frac{d\nu}{dm} = 2 m^2 P_{KP}(m)/{\widetilde{C}}(\nu)
\end{equation}
where ${\widetilde{C}}(\nu)$ denotes the inverted Laplace transform of $C(t)$.

Solution of the Eq.(\ref{f12}) allows us, in principle, to generate 
a stochastic time series of the process with a given, quite arbitrary 
covariance and with an arbitrary distribution $P_{KP}(m)$. 

\subsection{Exponential covariance}

The KP for the exponential covariance
\begin{equation}
\label{f13}
C(t)=\nu_0 \exp(-\nu_0 t),
\end{equation}
where $\nu_0=$const is a reciprocal of the correlation time, is called the 
Kubo-Anderson process \cite{kuan}. The jumping times are uniformly distributed 
in the interval $(-\infty,\infty)$ with a $m$-independent density $\nu_0$, 
according to the Poissonian distribution. Therefore the intervals of constant 
$m$ are distributed exponentially: $P(s)=\nu_0 \exp(-\nu_0 s)$. 
The value of the process $m(t)$ may be chosen 
according to an arbitrary distribution $P_{KP}(m)$. In that sense, 
the distributions of $m$ and $s$ are independent of each other. That property
holds only for the Kubo-Anderson process; for a general KP $m$ and $s$ are
interdependent. The Fokker-Planck equation (\ref{fpkp}) takes a simpler form
for the Kubo-Anderson process:
\begin{equation}
\label{fpka}
\frac{\partial}{\partial t}p(m,t) = \nu_0 \left(-p(m,t) + P_{KP}(m) 
\int p(m',t) dm'\right) .
\end{equation}

Some forms of $P_{KP}(m)$ are distinguished. The simplest choice is 
$P_{KP}(m)\sim \delta(m-\nu_0)+\delta(m+\nu_0)$ and corresponds to 
the dichotomous noise (the random telegraph process) \cite{gar}. 
Due to the central limit theorem, the 
Gaussian distribution of $P_{KP}(m)$ is of special importance. The
Kubo-Anderson process with that distribution resembles the Ornstein-Uhlenbeck
process. However, both processes are not identical \cite{doe}; the 
Fokker-Planck equation for the Ornstein-Uhlenbeck process,
$\frac{\partial}{\partial t}p(m,t) = \nu_0\frac{\partial}{\partial m}\left(m+
D\nu_0\frac{\partial}{\partial m}\right)p(m,t)$,
differs from the Eq.(\ref{fpka}). 

\subsection{Stretched exponential covariance}

The exponential distribution of interval lengths $P(s)$, a distinctive
feature of the Kubo-Anderson process, can also characterize stochastic
processes with the covariance form other than exponential. 
Let us consider the covariance possessing the shape known as the "stretched 
exponential". This function is the following:
\begin{equation}
\label{f14}
C(t) = \exp(-\alpha t^\gamma)
\end{equation}
where $\alpha=$const and $0<\gamma<1$. In the present study we assume 
$\gamma=1/2$. The stretched exponential function describes relaxation phenomena
in random systems \cite{kohwil,pal} and can be attributed to a 
dispersive transport of mobile defects in the glass \cite{shlkak}.
The dispersive transport is characterized by the infinite average time between 
subsequent hops. Moreover, the velocity autocorrelation function 
of a particle inside the Sinai billiard with the finite horizon 
is also given by the Eq.(\ref{f14}) \cite{bou1}.

We want to find expressions, useful in practical applications, allowing us 
to generate time series possessing the required covariance. 
The inverted Laplace transform of (\ref{f14}) reads
\begin{equation}
\label{f15}
{\widetilde{C}}(\nu) =\frac{1}{2} \frac{\alpha}{\sqrt{\pi}} \nu^{-3/2} 
\exp(-\alpha^2/(4\nu)).
\end{equation}
In order to simplify the differential equation (\ref{f12}),
we take the amplitude distribution in the form
\begin{equation}
\label{f16}
P_{KP}(m) = \frac{2}{\pi} |m|^{-5} \exp(-\pi^{-1} m^{-4}),
\end{equation}
where $m\in (-\infty,\infty)$. The distribution possesses maxima at 
$m=m_{max}=\pm [4/(5\pi)]^{1/4}$ and it is very small near $m=0$. 
Inserting Eq.(\ref{f15}) and Eq.(\ref{f16}) into the Eq.(\ref{f12}), 
we get the differential equation for $\nu(m)$ in the form
\begin{equation}
\label{f17}
\frac{d\nu}{dm} = 8/(\sqrt\pi \alpha)  m^{-3} \nu^{3/2} 
\exp[-\frac{\alpha^2}{4}(\nu^{-1}-4/(\alpha^2 \pi m^{4}))].
\end{equation}
One can easily check that Eq.(\ref{f17}) is satisfied by the function
\begin{equation}
\label{f18}
\nu(m)=1/s=(\pi \alpha^2/4)\;m^4
\end{equation}
and the interval lengths distribution is indeed exponential:
\begin{equation}
\label{f19}
P(s) =\frac{\alpha^2}{4} \exp(-\frac{\alpha^2}{4}s).
\end{equation}
The direct relation between $s$ and $m$ follows from the Eq.(\ref{f18}).
From that equation we conclude that $s\in (0,\infty)$ and long intervals
correspond to the values of $m$ close to zero; such events are 
extremely rare. Technically, a time series $m(t)$ can be constructed by
sampling subsequent intervals $s$ from the distribution $P(s)$. Then 
corresponding process values are evaluated by means of the Eq.(\ref{f18}), 
taking into account, in addition, that both signs of $m$ are equally probable. 

\subsection{Algebraic covariance}

Let us now consider the KP possessing the power-law covariance that we express
in the following form
\begin{equation}
\label{f20}
C(t) = \Gamma(\gamma)\; t^{-\gamma}~~~~~~(\gamma>0).
\end{equation}
The jumping frequency $\nu(m)$ one can derive from Eq.(\ref{f12}), similarly as
for the stretched exponential case:
\begin{equation}
\label{f21}
\nu(m) = \left[2\gamma\int_0^{|m|} m'^2 P_{KP}(m')dm'\right]^{1/\gamma}.
\end{equation}
We assume the amplitude distribution in the algebraic form: 
$P_{KP}(m)\sim m^{-\alpha}~~(\alpha=\mbox{const})$, modified in order 
to satisfy the normalization condition $2\int_0^\infty P_{KP}(m)dm=1$. 
Generally, two different forms of $P_{KP}(m)$ are possible.

In the first case we cut off the large values of $|m|$:
\begin{eqnarray}
\label{f22}
P_{KP}(m)=\cases {\frac{1-\alpha}{2a}(|m|/a)^{-\alpha}&$|m|\le a$\cr
0&$|m|>a$}
\end{eqnarray}
where $a=$const is an additional parameter. Due to the
condition $\nu(\infty) = \infty$, $a$ must be a large number; finiteness of $a$
results in a deviation of the covariance from the assumed form (\ref{f20}) 
near $t=0$ and in removal of the singularity. Moreover, 
the normalization condition implies $\alpha<1$. Inserting $P_{KP}(m)$ to the
Eq.(\ref{f21}) gives us the expression for the relation between $m$ and $s$:
\begin{equation}
\label{f23}
s=\frac{1}{\nu}=(\gamma\alpha')^{-1/\gamma} 
a^{(1-\alpha)/\gamma} |m|^{-(3-\alpha)/\gamma} \theta(s-\epsilon)
\end{equation}
where we have introduced a constant $\alpha'=(1-\alpha)/(3-\alpha)$.
Finiteness of the parameter $a$ imposes a restriction on the 
lower bound of the interval length $\epsilon$: $s\in(\epsilon,\infty)$. 
The smallest interval length $\epsilon$ is related to that parameter by: 
$\epsilon=(\alpha'\gamma)^{-1/\gamma}\;a^{-2/\gamma}$.
The probability distribution of interval lengths takes the algebraic form
\begin{equation}
\label{f24}
P(s)=(\gamma\alpha')^{1-\alpha'} a^{-2\alpha'} s^{-\gamma\alpha'-1}.
\end{equation}
From the Eq.(\ref{f24}) some restrictions on possible asymptotic behaviour of
$P(s)$ follow. The slowest decay rate for large $s$ occurs for $\alpha$ close 
to 1: $P(s) \sim s^{-1}$. On the other hand, the distribution $P(s)$ falls 
the most rapidly, as $\sim s^{-\gamma-1}$, for $\alpha\to -\infty$.

The other possibility is to cut off the distribution $P_{KP}(m)$ near $m=0$:
\begin{eqnarray}
\label{f25}
P_{KP}(m)=\cases {0&$|m|<a$\cr
\frac{\alpha-1}{2a}(|m|/a)^{-\alpha}&$|m|\ge a$}
\end{eqnarray}
where $\alpha>1$. One can show that the minimal interval length is
finite (non-zero) if $\alpha>3$. The interval lengths distribution can be 
obtained similarly as for the case (\ref{f22}). 
The result is slightly more complicated: 
\begin{equation}
\label{f26}
P(s)=\left(\frac{3-\alpha}{\gamma(\alpha-1)}a^{1-\alpha} s^{-\gamma}+
a^{3-\alpha}\right)^{-2/(3-\alpha)} s^{-\gamma-1}.
\end{equation}
Asymptotically the distribution approaches $s^{-\gamma-1}$, independently of
$\alpha$.

\section{Application to the GLE}

We will now consider the Eq.(\ref{gle}) for which the random force $F(t)$ is
modeled by means of a concrete stochastic process, possessing a given 
covariance, and simulate stochastic trajectories of the Brownian particle 
by means of a Monte Carlo method. Therefore, two equations are to be solved 
simultaneously: the original GLE and the second one, describing an adjoined 
process, in the form of some KP. Accordingly, in the following we assume 
$F(t)=m(t)$. A similar approach used to be applied \cite{jung,zhi} to deal 
with stochastic equations driven by the coloured noise; the adjoined process 
constitutes in that case the Ornstein-Uhlenbeck process.
Independently of those methods, which model the noise as a stochastic, random
process, one can introduce some deterministic system possessing a required
autocorrelation function. Value of the noise at a given time is then 
determined by the evolution of dynamical equations of motion. In this way
Shimizu \cite{shi} solved both ordinary Langevin equation and GLE, 
representing the noise by a chaotic map. Similarly, the Sinai billiard has been
used to model a noise with long-time covariance 
in the Langevin equation \cite{sin}.
Certainly, the first two moments do not determine the noise uniquely and
the choice of a model involves additional assumptions, especially for 
non-Gaussian and not exponentially correlated processes. The form of the noise
must be decided according to physical requirements of a concrete application.
Nevertheless, some important quantities, like the autocorrelation function of
the Brownian particle velocity (\ref{codt}), apparently depend only on 
the noise covariance. 

However, for some stochastic processes description of those quantities provided
by the equations collected in the Introduction does not agree with the
simulation results. 
Solving the GLE to obtain the Brownian particle velocity requires the value of
the noise $F(t)$, determined by the adjoined process, at a given time. It has
been demonstrated recently \cite{para} that this requirement modifies
probability distributions -- the adjoined process looks differently when
inserted into the equation. Consequences of that change for the Monte Carlo 
simulation results may be important. In particular, one can expect a violation
of the FDT which manifests itself in an improper asymptotic behaviour 
of the velocity variance $\langle v^2(t)\rangle_{S\;}$. 
Determining of that quantity can indicate whether the expected equilibrium 
state is reached and the equipartition energy rule holds. 

The Brownian particle velocity is to be determined from the stochastic integral
(\ref{solv}) in which step-wise, constant values of $F(t)$, $F_k$, follow 
from the length of subsequent time intervals $s_k$:
\begin{equation}
\label{solv1}
v(t)=m^{-1}\left[\sum_{k=1}^n F_k\int_{t_{k-1}}^{t_k} R(t-\tau)\,d\tau
+F_{n+1}\int_{t_n}^t R(t-\tau)\,d\tau\right],
\end{equation}
where $t_k-t_{k-1}=s_k$ and $t_0=0$. Sampling of intervals $s_k$ continues as
long as the time $t$ is reached: $\sum_1^n s_k < t$ and $\sum_1^{n+1}s_k\ge t$,
where $n$ is an arbitrary integer. The interval length distribution $P(s)$ 
is a natural quantity of interest. The last, $n+1$, interval corresponds to 
the time $t$ \cite{uwa2}. It is clear that the
distribution of lengths of that interval, 
the "effective" interval distribution $\widehat{P}(s,t)$, cannot be 
identical with $P(s)$; a simple consideration reveals e.g. the enhanced 
probability of choosing longer intervals. Generally, that modified 
distribution can depend on $t$. One can express it in terms of the cumulative 
distribution function $\Phi(s,t)$ \cite{para}: 
\begin{equation}
\label{phat}
\widehat{P}(s,t)=\frac{\partial}{\partial s} \Phi(s,t),
\end{equation}
where
\begin{eqnarray}
\label{phi}
\Phi(s,t) = \cases {\int_{t-s}^t S(x) dx \int_{t-x}^s P(\xi) d\xi
& for $~~~0\le s\le t$\cr
\int_0^t S(x) dx \int_{t-x}^s P(\xi) d\xi + 
\int_t^s P(\xi) d\xi& for $~~~~~~~~~\!s>t.$}
\end{eqnarray}
The form of auxiliary function $S(x)$ follows from the normalization condition
\begin{equation}
\label{sodx}
\int_0^t S(x) dx \int_{t-x}^\infty P(\xi) d\xi + \int_t^\infty P(\xi) d\xi= 1.
\end{equation}

For the Kubo-Anderson process, the modification of the interval distribution
$P(s)$, exponential in that case, is of minor importance because $P_{KP}(F)$ is
independent of $P(s)$ and the interval lengths do not influence process 
values. Consequently, results of the simulations agree with
general predictions implied by the FDT. Other forms of noise covariance 
require taking into account the modified distribution $\widehat P(s,t)$. 
The case of stretched exponential covariance provides a simple but non-trivial 
example. It involves also the Poissonian distribution of jumping times.

\subsection{Stretched exponential noise covariance}

We assume the noise covariance in the following form:
\begin{equation}
\label{csex}
C_F(t) = 2mT/\alpha^2 \exp(-\alpha \sqrt t).
\end{equation}
The Eqs.(\ref{solv}) and (\ref{solv1}) express 
the solution of GLE in terms of the 
resolvent $R(t)$. The Laplace transform of that function, $\widetilde R(s)$, 
is given by the Eq.(\ref{rods}) and kernel has the form:
\begin{equation}
\label{ksex}
\widetilde K(s) = \frac{\alpha^2}{2} \left(\frac{1}{s} - 
\frac{\sqrt\pi\alpha}{2s\sqrt s} \exp(\alpha^2/(4s)) 
(1-\mbox{erf}(\frac{\alpha}{2\sqrt s}))\right)
\end{equation}
where erf$(x)$ denotes the error function. To obtain the resolvent $R(t)$ we
need to evaluate the inverse Laplace transform from $\widetilde R(s)$: 
$R(t)=1/2\pi i\; \int_{-i\infty +\sigma}^{+i\infty +\sigma}\widetilde R(z)\,
\mbox{e}^{\,tz} \;dz$. The integrand possesses 
two conjugate simple poles and a cut along the negative real axis.
Evaluation of the contour integral produces the following result
\begin{equation}
\label{rodt}
R(t)=\mbox{e}^{-at}\,(c_1\sin bt +c_2\cos bt)-
{\displaystyle \frac{4}{\sqrt\pi} \int_0^\infty \frac{x^2
\exp[x^2-\alpha^2t/(4x^2)]\;dx}
{[(2x^2+\alpha^2/(4x^2)) \exp(x^2)-2\sqrt\pi x^3 
\mbox{erfi}(x)]^2+4\pi x^6}}~~;
\end{equation}
the imaginary error function erfi $(x)\equiv -i$ erf $(ix)$ can easily be
calculated by the following expansion:
\begin{eqnarray}
\label{erfi}
\mbox{erfi}(x)=\frac{2}{\sqrt\pi} \sum_{n=0}^\infty 
\frac{x^{2n+1}}{n!(2n+1)}  \nonumber ~~.
\end{eqnarray}
The constants $a$ and $b$ denote the real and imaginary parts of the pole of
$\widetilde R(z)$, respectively: $z_0=-a-|b|i$; they have to be evaluated
numerically. $c_1$ and $c_2$ can be found by the standard residues analysis.
For $\alpha=1$ the constants are the following: $a=0.207094$, $b=0.440963$,
$c_1=-0.127752$, and $c_2=0.593952$ . Fig.1 presents the function $R(t)$ for
$\alpha=1$ and $\alpha=2$. 

We wish to perform the Monte Carlo simulation using the noise defined by 
the process (\ref{f16}). According to
(\ref{f19}), the interval distribution is exponential: 
$P(s)=\beta\exp(-\beta s)$, where $\beta=\alpha^2/4$. We expect that taking
into account of the modified form of the distribution may be important for the
simulation results because $m$ and $s$ are connected. That distribution can 
easily be found in this case. From the Eq.(\ref{sodx}) we obtain $S(x)=\beta$; 
finally we get:
\begin{eqnarray}
\label{pnex}
{\widehat P}(s,t)=\cases {\beta^2 s \exp(-\beta s)
& for $~~~0\le s\le t$\cr
\beta(1+\beta t)\exp(-\beta s)& for $~~~~~~~~~\!s>t.$}
\end{eqnarray}
The function (\ref{pnex}) is presented in Fig.2.
The distribution is discontinues. It depends on $t$ but this dependence 
dwindles exponentially with time; the left-hand branch ($s\le t$) is
time-independent. Therefore, asymptotically the process becomes 
stationary. Nevertheless, ${\widehat P}(s,t)$ possesses the mean value 
twice of that for $P(s)$ \cite{uwa1} and simulation results must reflect that.
Indeed, the velocity variance obtained from the Eq.(\ref{solv1}), shown in 
Fig.3, stabilizes at a lower value then that predicted by the equipartition 
energy rule ($T/m$). It is so because long intervals correspond to small values
of the noise amplitude, according to the Eq.(\ref{f18}). For short times, in
turn, the result of the simulation agrees with the general prediction 
(\ref{v2a}), also shown in the Fig.3, because then the branch $s>t$ dominates 
the distribution (\ref{pnex}) and the dependence on $t$ is weak. 

It is possible to construct some KP which does not change when inserted into
the GLE and which produces the equilibrium state in agreement with the FDT. 
For that purpose we single out some subset of kangaroo processes, 
a {\it restricted} KP (RKP), defined in the following way. We choose subsequent
intervals $s_k$ in the step-wise evolution of KP, according to the distribution
$P(s)$, and assume that $t$ corresponds to $n+1$ interval, i.e. 
$S_n\equiv s_1+s_2+\dots +s_n<t$ and $S_{n+1}>t$. Let $S_t\equiv S_{n+1}-t>0$.
We call some KP "restricted" if $S_t\le d$ for a given $d$; 
$n$ is an arbitrary integer: $n\in (0,\infty)$. The probability distribution of
the last interval lengths we denote by $P^\star(s,t;d)$. Obviously, 
$P^\star(s,t;\infty)={\widehat P}(s,t)$. On the other hand, in the limit
$d\rightarrow 0$ the last interval lengths obey the original, 
time-independent distribution $P(s)$ but only for $s\le t$ because 
longer intervals are excluded by construction. Therefore the distribution
$P(s)$ can be recovered at large times if those intervals are negligible. 
We get then the following theorem: If the probability that intervals 
in a sequence $s_k$ are larger than $t$ asymptotically vanishes, then
\begin{equation}
\label{twier}
\lim_{t\rightarrow\infty}\lim_{d\rightarrow 0} P^\star(s,t;d)=P(s).
\end{equation}

For the distribution (\ref{pnex}) the required probability vanishes 
with time and the theorem can be applied. In practice it
can be done easily by choosing some small $d$ and sampling intervals from the
distribution $P(s)$, in the same way as before, as long as the time $t$ 
is reached. Then all sequences of intervals for which $S_t>d$ are rejected. 
Fig.3 presents the result of such calculation for $d=0.01$. At short times, 
the velocity variance differs substantially from the other results shown in the
Figure because the branch $s>t$ is then essential but asymptotically it 
approaches the value $T/m$, in accordance with the equipartition energy rule. 

\subsection{Algebraic noise covariance}

Finally, we consider a power-law form of the covariance: 
\begin{equation}
\label{cpl}
C_F(t)\sim t^{-\gamma}~~~~~~~~~~(0<\gamma<3) .
\end{equation}
The KP we apply to model the noise is defined by Eq.(\ref{f22}); we implement 
the simplest case $\alpha=0$. Then the amplitude distribution is a constant
(except very large $|F|$): 
$P_{KP}(F)=\sqrt{\gamma\epsilon^\gamma/3}/2$ where $\epsilon$ is the smallest
interval length. The interval distribution follows from Eq.(\ref{f24}):
\begin{equation}
\label{pods}
P(s)=\frac{\gamma}{3} \epsilon^{\gamma/3} s^{-1-\gamma/3}\;\theta(s-\epsilon) .
\end{equation}

The resolvent $R(t)$ for the noise with covariance (\ref{cpl}) can be evaluated
by means of similar methods as for the stretched exponential covariance.
Results can be found in Ref.\cite{kubo1} for $\gamma=3/2$ and in 
Refs.\cite{para,glang} for $\gamma=1$. Now we want to calculate the modified 
distribution ${\widehat P}(s,t)$. First, we have to solve the Eq.(\ref{sodx}) 
which assumes the form of Abel's integral equation
\begin{equation}
\label{abel}
\int_0^t S(x)(t-x)^{-\gamma/3}dx+t^{-\gamma/3}=\epsilon^{-\gamma/3}.
\end{equation}
The solution reads
\begin{equation}
\label{sols}
S(x)=\frac{\epsilon^{-\gamma/3}}{\Gamma(1-\gamma/3)\Gamma(\gamma/3)}
\;x^{\gamma/3-1}-\delta(x).
\end{equation}
After evaluation integrals, we obtain from Eqs.(\ref{phat}) and (\ref{phi}) the
expression for the required distribution:
\begin{eqnarray}
\label{pnods}
{\widehat P}(s,t)=\cases {{\displaystyle\frac{s^{-\gamma/3-1}}
{\Gamma(1-\gamma/3)\Gamma(\gamma/3)}} 
\left[t^{\gamma/3}-(t-s)^{\gamma/3}\right]
& for $~~~\epsilon\le s\le t$\cr
\cr
{\displaystyle\left[\frac{t^{\gamma/3}}{\Gamma(1-\gamma/3)\Gamma(\gamma/3)}+
\frac{\gamma\epsilon^{\gamma/3}}{3}\right]}\;s^{-\gamma/3-1}& 
for $~~~~~~~~~\!s>t.$}
\end{eqnarray}
The distribution ${\widehat P}(s,t)$ for $\gamma=1$ is presented in Fig.4. 
The picture is markedly different from that obtained for the exponential case 
(Fig.2); the right branch, corresponding to the intervals $s>t$, does not
vanish with time but gets larger comparing to the left branch. The entire 
distribution shifts with time towards large intervals because the average
interval length is infinite. 

Therefore the effective distribution ${\widehat P}(s,t)$ differs substantially
from the $P(s)$, and the stochastic process
generated by it must possess different properties. First, let us recalculate 
the covariance that, in general, can depend on an initial time $t_0$: 
${\widehat C}(t,t_0)=\langle F(t_0)F(t_0+t)\rangle_{S\;}$, where the process
$F(t)$ is to be determined by the simulation. Technically that means that 
for a given $t_0$, one produces a sequence of intervals to reach the time
$t_0+t$. Then one evaluates corresponding values of the process $F(t)$ 
using Eq.(\ref{f23}). Then the process is governed by the distribution 
${\widehat P}(s,t)$. The expression for the covariance follows from 
the Eq.(\ref{f10}):
\begin{equation}
\label{cnkp}
{\widehat C}(t,t_0) = 
\int_\epsilon^\infty s^{-2\gamma/3} \exp(-t/s) {\widehat P}(s,t_0)ds.
\end{equation}
Evaluation of the integral gives the following result:
\begin{eqnarray}
\label{cnkp1}
\lefteqn{{\widehat C}(t,t_0) = \frac{3\epsilon^{-\gamma/3}\gamma^{-1/3}}
{\Gamma(1-\gamma/3)\Gamma(\gamma/3)}
[t_0^{\gamma/3} t^{-\gamma}{\bar\gamma}(\gamma,t/\epsilon)} \nonumber \\
 & \\
&~~~~~~~~~ -t_0^{1/2-\gamma/6} t^{(-1-\gamma)/2}\exp(-t/2t_0)
\Gamma(1+\gamma/3)\;
W_{\gamma/6-1/2,-\gamma/2}\;(t/t_0)], \nonumber
\end{eqnarray}
%
where ${\bar\gamma}(a,x)$ denotes the incomplete gamma function and 
$W_{\alpha,\beta}\;(x)$ stands for the Whittaker function \cite{wit}. 
Certainly, the above result is different from our starting covariance
(\ref{cpl}); the most striking feature of the function ${\widehat C}(t,t_0)$ 
is the dependence on $t_0$ which does not diminish with $t$. The variance of 
the process, ${\widehat \sigma}^2(t_0)$, 
can be found by inserting $t=0$ into the Eq.(\ref{cnkp1}). Let us consider two
examples. The case $\gamma=1$ has been discussed in Ref.\cite{para}; 
the final expression for the variance is the following: 
\begin{equation}
\label{f2n}
{\widehat \sigma}^2(t_0)=\frac{\epsilon^{-1/3}}{\Gamma(1/3)\Gamma(2/3)}\;
\left[3\ln3/2+\pi\sqrt{3}/6+\ln(t_0/\epsilon)\right]\;
t_0^{-2/3}~~~~~~~~~~(t_0\gg\epsilon).
\end{equation}
Algebraically correlated stochastic processes for $\gamma=3/2$ are especially
important. They have been extensively studied in connection with the 
Brownian motion in a viscous fluid \cite{bouss,maz,cho,ald,kubo1}. In
this case the variance of the process reads
\begin{equation}
\label{f2nh}
{\widehat \sigma}^2(t_0)=(\frac{2}{3})^{1/3}\left(\frac{2}{\pi\epsilon}\;
t_0^{-1/2}+t_0^{-3/2}\right).
\end{equation}
Therefore, in contrast to the original variance (calculated with the
distribution $P(s)$) $\sigma^2=\gamma^{-1}\epsilon^{-\gamma}$, the effective 
variance ${\widehat \sigma}^2$ is time-dependent and tends to zero. That
behaviour is a direct consequence of nonstationarity, i.e. of the 
time-dependence of the distribution ${\widehat P}(s,t)$. The decline of
${\widehat \sigma}^2(t_0)$ means that the effective temperature of the system
drops to zero when we insert the process into the stochastic integral 
(\ref{solv1}) as a model of the noise. Consequently, the
Brownian particle velocity variance also must dwindle with time. Indeed,
a direct Monte Carlo simulation confirms this conclusion, as it has been
demonstrated in Ref.\cite{para} for $\gamma=1$.

The RKP can be constructed also for algebraic correlations. However, the
theorem (\ref{twier}) cannot be applied because of the strong time-dependence 
of the distribution ${\widehat P}(s,t)$. The probability that the interval
length is larger then $t$ does not decline with time: 
${\widehat P}(s>t,t)=\int_t^\infty {\widehat P}(s,t)ds=
3/[\gamma\Gamma(1-\gamma/3)\Gamma(\gamma/3)]=$const.

\section{Summary and discussion}

The kangaroo processes represent a broad class of random functions
characterized by various forms of the covariance. Therefore, they provide 
an opportunity to model physical stochastic processes possessing an arbitrary 
covariance and, in addition, a quite general amplitude distribution. The KP is
step-wise -- the value of the process changes according to some jumping
frequency which, in turn, depends on that value. 
Some physical phenomena exhibit a similar, step-wise behaviour and the KP 
is a natural process to model them. An important quantity is the
distribution of intervals of constant process value $P(s)$, uniquely connected
to the amplitude distribution. 
In this paper, we have discussed two forms of that distribution: the
exponential and the algebraic ones. We have demonstrated how one can generate
algebraically correlated processes using the KP with some algebraic $P(s)$. The
exponential form, in turn, is suitable to represent both exponential and 
stretched exponential correlations. 

The GLE has been solved using KP as a model of the noise. The problem has
been considered as a juxtaposition of two random processes: of velocity of the 
Brownian particle, described by the GLE, and of the adjoined KP. In the 
framework of that approach, the force $F(t)$ in the stochastic integral 
(\ref{solv}) is determined by looking for a value of the KP, 
{\it independently} evolved, at a given time. 
Such procedure changes probability distributions 
of the KP and Monte Carlo simulated solutions of the GLE are not
in agreement with results predicted by general analysis, founded on the FDT. 
Results obtained from effective probability distributions, 
among which the interval distribution ${\widehat P}(s,t)$ 
is the most important, does not correspond to the equilibrium state 
consistent with the equipartition energy rule.
For algebraically correlated processes, even nonstationarity effects emerge. 
Are those results involving effective distributions a necessary consequence 
of modeling of the noise by means of adjoined random process in the form of 
KP? If the stochastic force we want to insert into the stochastic integral 
(\ref{solv}) represents some independent physical process, the modified 
probability distribution (\ref{phat}) has to be taken into account. 
In such case we must expect an apparent violation of the FDT, despite proper
definition of the kernel. Reversely, any information about
properties of the noise, extracted from GLE solutions, always refers to the
effective appearance of that process in the GLE. 

On the other hand, for some cases there is a possibility to construct
the model in such a way to avoid any modification of the distributions 
and to preserve consistency with the FDT. For that purpose, one can use 
some specific version of the ordinary KP -- the RKP. 
The idea is very simple: if some interval in the
step-wise evolution of the KP ends exactly at a point corresponding to the
required time, nothing has to be modified. We can imagine the KP as a "clock"
with a variable frequency given by the distribution $P(s)$. In that picture,
applying of the RKP with $d=0$ means a synchronization of that clock in respect
to the physical time in the GLE. An important limitation of the 
synchronization procedure consists in the fact that interval lengths of the 
RKP are always finite, not larger than the time $t$, and $P(s)$ usually 
possesses infinite tails. If, however, those tails decline sufficiently fast, 
intervals larger then $t$ become negligible. It has been demonstrated that the
exponential interval distribution possesses this property and a Monte Carlo
algorithm utilizing the RKP can easily be constructed -- simulation results
indeed correspond to the equilibrium state predicted by the FDT. 
The KP in this form is especially suitable for physical applications. 
We note, however, that the RKP with $d=0$ is not completely 
independent of the GLE: the synchronization introduces a coupling.
The case of stretched exponential covariance is only the simplest non-trivial 
example of application of KP with the Poissonian interval lengths 
distribution. A straightforward generalization, allowing for the other 
power-law dependences $m(s)$, produces KP's with covariances given by 
the Bessel functions K$_\nu(\sqrt t)$. 

One cannot expect that every shape of the covariance function may be modeled 
by some sufficiently steep form of $P(s)$ and, therefore, the RKP is always 
a proper tool. Processes possessing covariances with long tails are
characterized by long intervals (long free paths). Indeed, in Sec.II we have 
indicated a strong limitation of admissible shapes of $P(s)$. The algebraic 
tails of $P(s)$ are essential and very long intervals given by the effective 
distribution ${\widehat P}(s,t)$ are not negligible for large times; the
distribution itself remains time-dependent. For that reason, 
the RKP which consist in cut off of long intervals, 
cannot work as a model of algebraic covariances. In the other words: due to
divergence of moments of both distributions, $P(s)$ and ${\widehat P}(s,t)$, 
the branch corresponding to the intervals $s>t$ is important for large $t$ and 
the synchronization cannot be achieved.

Necessity of taking into account the modified form of probability 
distributions is not restricted to the GLE 
but relates to any stochastic equation, e.g. the
ordinary Langevin equation, if the random force is modeled by the adjoined KP. 
Monte Carlo simulations can be useful for some generalizations 
of the diffusion equation, in particular for the Burgers equation \cite{bur}. 
That nonlinear equation, possessing a broad spectrum of applications in the
fluid dynamics, is characterized by long range noise correlations both in space
and time \cite{for,med}. Also that noise can possibly be modeled by the KP.

{\bf Acknowledgements}

The work was partly supported by KBN Grant No. 2 P03 B 07218.

\pagebreak

{\bf Figure captions}

FIG. 1. The resolvent $R(t)$ (\ref{rodt}) for the stretched exponential shape 
of noise covariance as a function of time. 

FIG. 2. The time evolution of the effective interval distribution 
${\widehat P}(s,t)$ (\ref{pnex}) corresponding to the exponential form of 
the original distribution $P(s)$ with $\beta=1$.

FIG. 3. The velocity variance obtained from the Monte Carlo simulation of GLE
solutions (\ref{solv1}) for stretched exponential form of noise covariance. 
The noise has been modeled by the KP (solid line) and the RKP with $d=0.01$, 
using theorem (\ref{twier}) (dashed line). 1000 trajectories have been
calculated for either case. The variance calculated from Eq.(\ref{v2a}) is
also presented (dots). The temperature $T=1$, the particle mass $m=1$, 
and the parameter $\alpha=1$.

FIG. 4. The time evolution of the effective interval distribution 
${\widehat P}(s,t)$ (\ref{pnods}) corresponding to the original distribution
$P(s)$ in the form (\ref{pods}) with $\gamma=1$. The minimal 
interval length $\epsilon=0.01$. 


\begin{thebibliography}{99}
\bibitem{uwa2}
Since the resolvent $R(t)$ has maximum at $t=0$, the value of KP corresponding 
to the last interval, $F_{n+1}$, enters the Eq.(\ref{solv1}) with 
the largest weight.
\end{thebibliography}
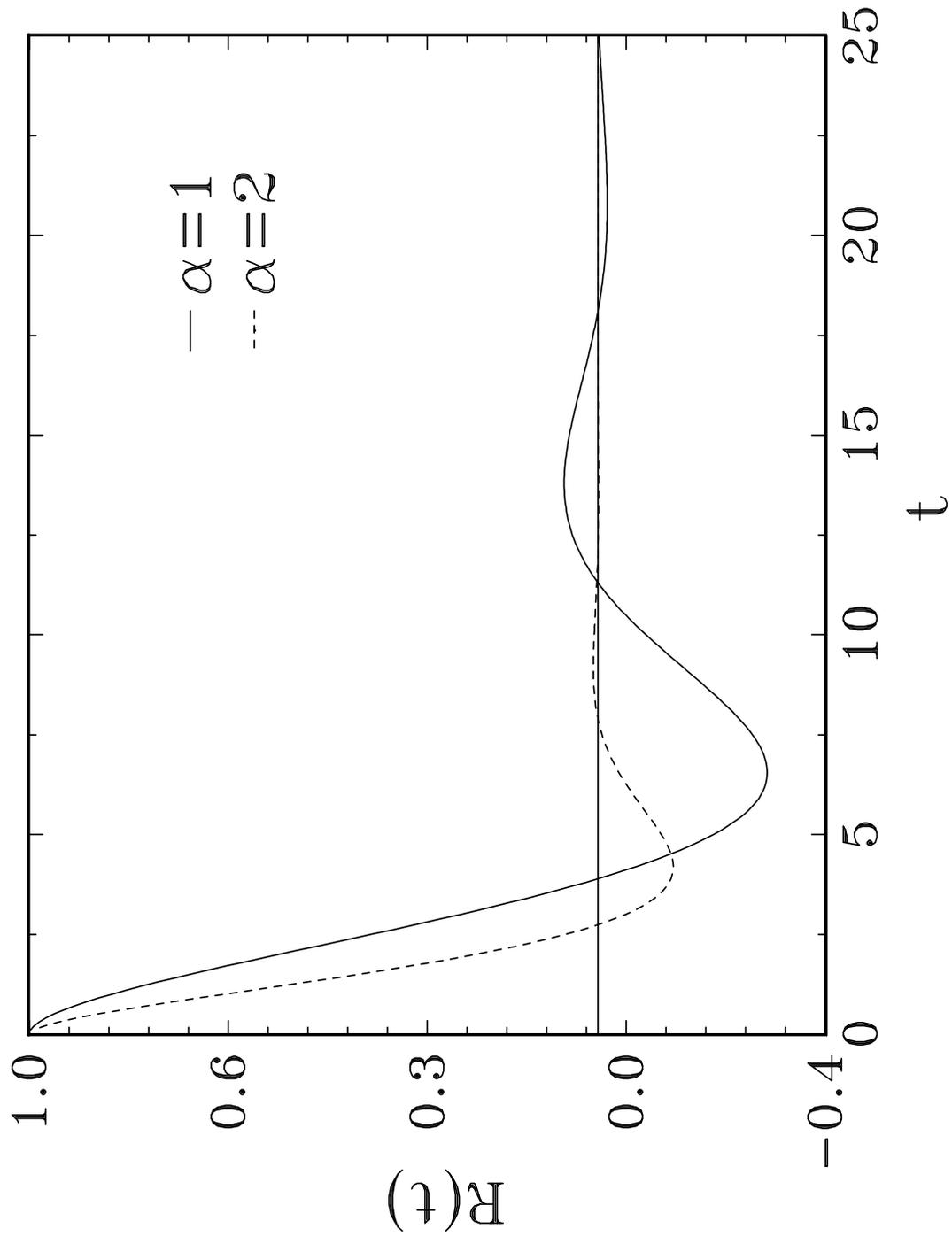
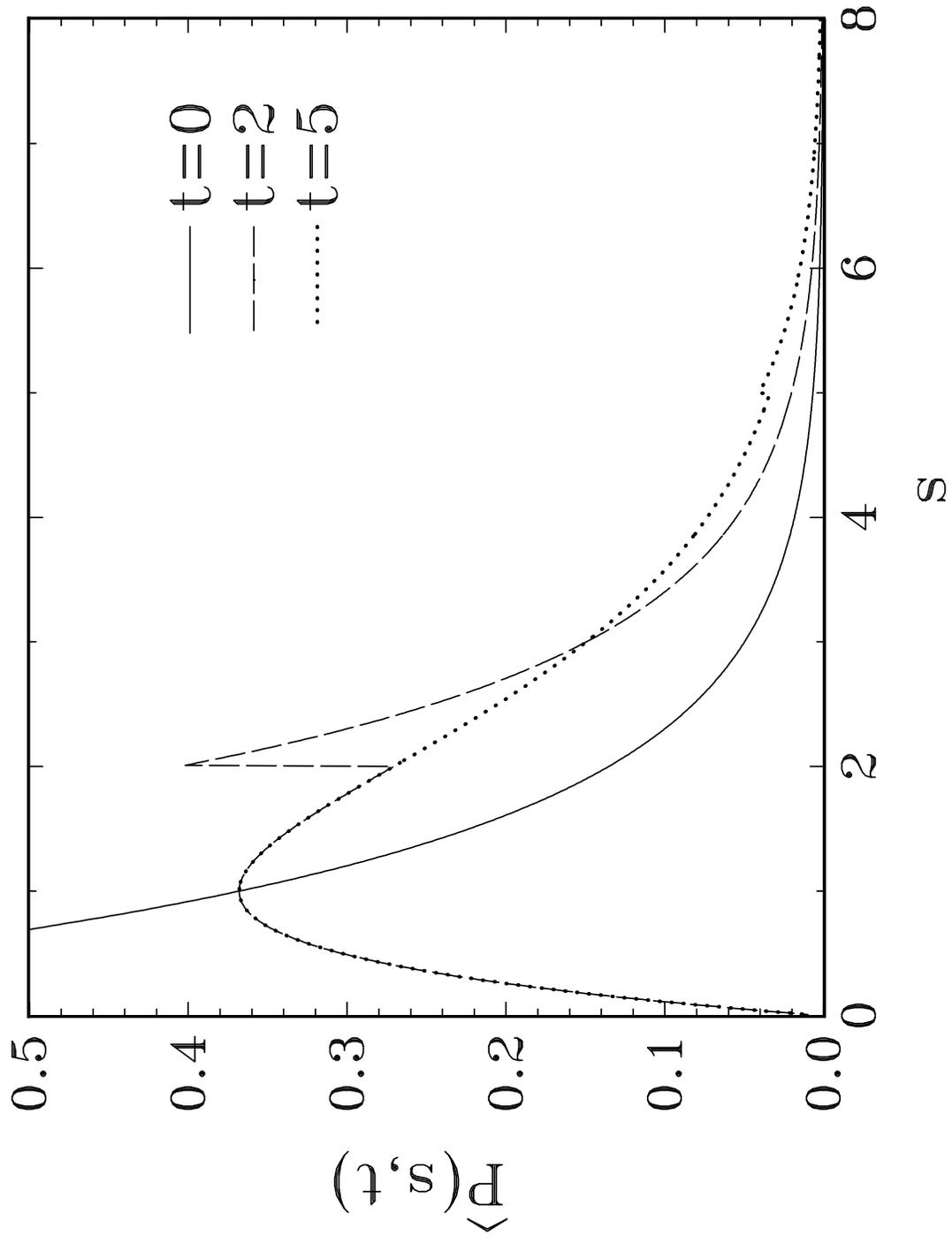
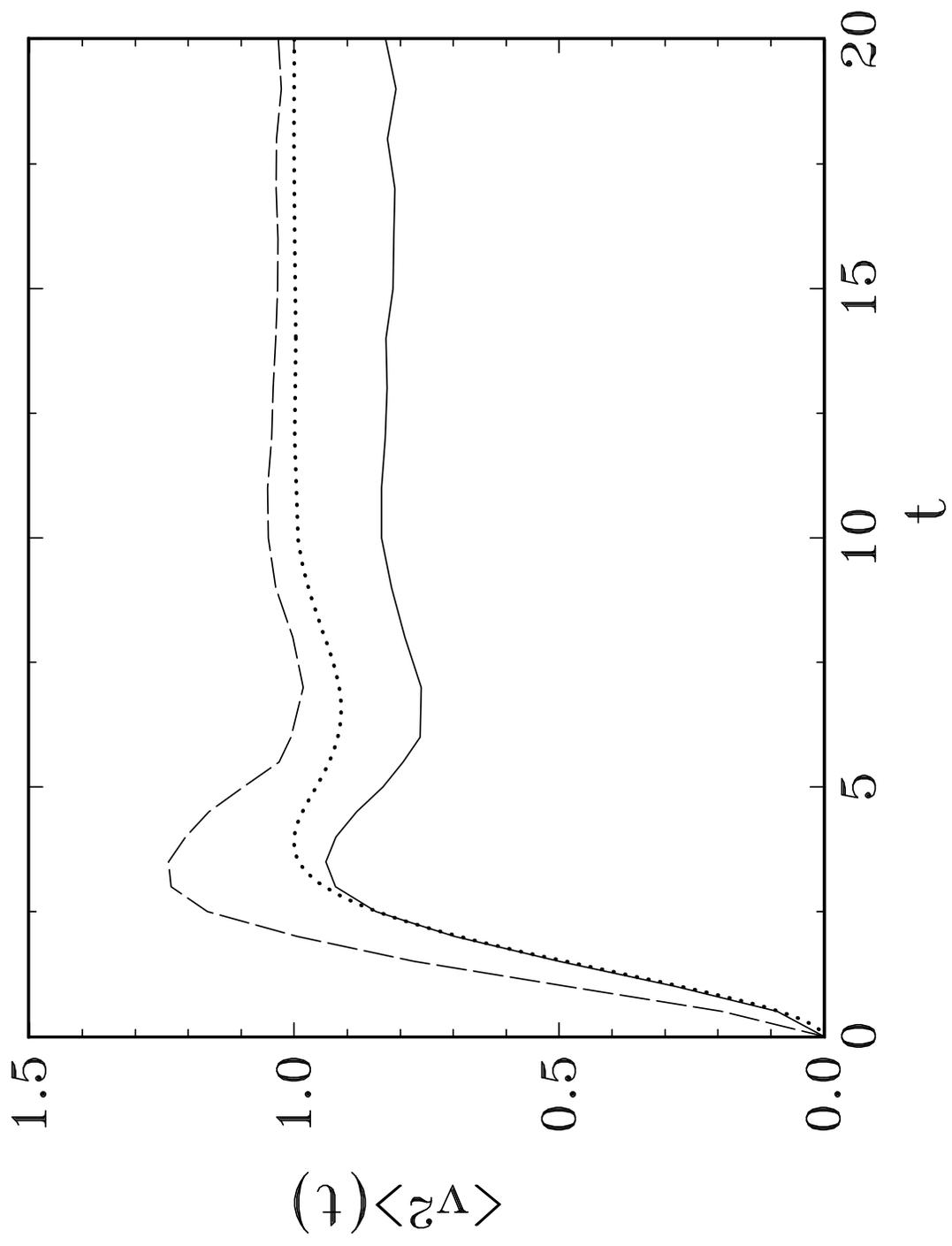
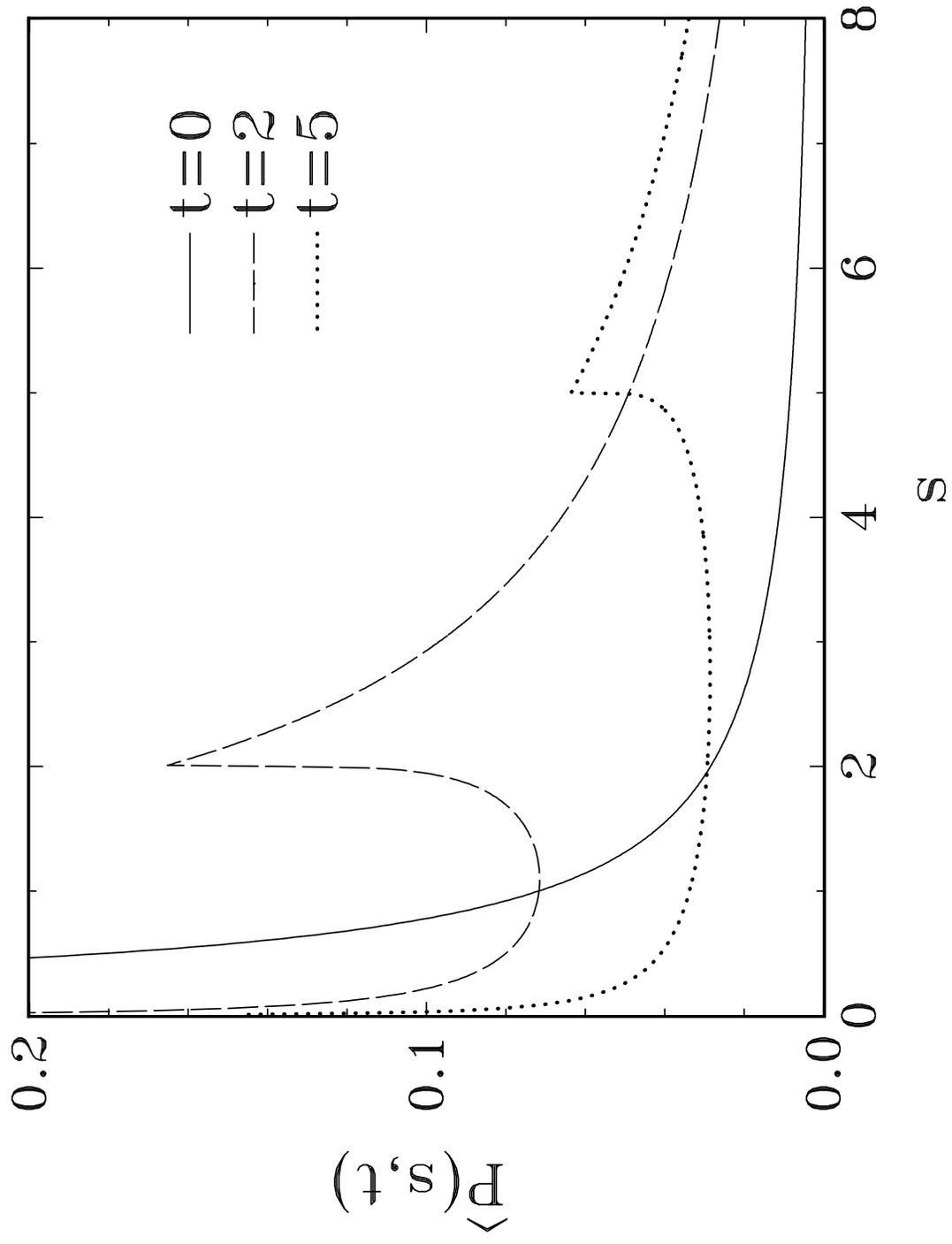
\end{document}